# Broadband Photo- and Electroluminescence from Bulk Silicon via Strong Photonic Confinement

*Aleksei I. Noskov[1], Alexander B. Kotlyar[2], Eric O. Potma[1], Dmitry A. Fishman[1]*

*[1]Department of Chemistry, University of California, Irvine, Irvine, CA 92697, USA*

*[2]George S. Wise Faculty of Life Sciences, Tel Aviv University, Tel Aviv 6997801, Israel*

## Abstract

Silicon's indirect bandgap fundamentally limits its ability to emit light, hindering the development of silicon-based light sources. Here, we explore a conceptually new solution to this long-standing challenge. We demonstrate ultrabroadband photo- and electroluminescence from bulk silicon, enabled by a radiative pathway mediated by momentum-expanded photonic states that bypass phonon-assisted transitions. This mechanism, previously demonstrated using metallic nanoparticles as photon confiners, is here realized in an all-silicon system using embedded sub-1.5 nm silicon nanoparticles. Since such ultrasmall particles possess negligible intrinsic emission efficiency, we instead demonstrate that they act as photonic confiners, enabling radiative recombination in the surrounding bulk material. The agreement with prior metal-based systems confirms that confinement size, rather than material composition, governs the activation of radiative transitions in a momentum-forbidden system. The emission spans the visible to near-infrared spectral range, with electroluminescence in an undoped semiconductor device visible under ambient conditions and a quantum efficiency estimated as ~0.2%. These findings establish a new route to efficient light emission in silicon and reveal a hybrid light-matter regime in which extreme photonic confinement reshapes the electronic transition landscape.

It is hard to overstate the technological importance of silicon. As the foundational material of modern electronics, it supports countless platforms on which our society depends. Yet despite its abundance, exceptional electrical properties, and highly developed processing infrastructure, silicon suffers from a well-known shortcoming: its indirect bandgap, which renders efficient light emission fundamentally impractical. Overcoming this limitation, while preserving silicon's compatibility with large-scale manufacturing, holds transformative potential.[1] If achieved, it could enable integrated silicon photonics, novel light sources, and entirely new paradigms in optoelectronic and photonic computing.

A wide range of strategies has been explored over the decades to address silicon's intrinsically low radiative efficiency. Among the most notable are efforts based on quantum confinement in nanoscale silicon structures - a direction shaped by the visionary works of Canham[2], Brus[3,4], Hybertsen[5], and Pavesi.[6-8] In these highly confined systems, known as quantum dots, enhanced spatial and momentum-space overlap of electron and hole wavefunctions promotes more efficient radiative recombination, even in an indirect bandgap material.[9-15] When confined to dimensions ~3 nm, silicon exhibits narrowband emission spectra characteristic of quantum dots, with peak wavelengths that strongly depend on particle size.[16,17] Other approaches have pursued fundamentally different paths to enhance the light–matter interactions in silicon. These include the use of subwavelength resonators, either as isolated nanostructures or periodic metasurfaces, leveraging Auger-assisted generation of hot electrons or Purcell enhancement of radiative rates.[18-20] Another set of works has focused on boosting silicon emission by increasing the local optical density of states through coupling with localized surface plasmon resonances.[21-24]

In this study, we introduce a conceptually new approach. It builds on our recent demonstration of broadband photoluminescence from bulk silicon wafers decorated with sub-1.5 nm gold or copper nanoparticles.[25] The spatial confinement of light is believed to be the origin of this bright emission, through expansion of the accessible photonic momentum states that enable phonon-independent radiative transitions in an otherwise momentum-forbidden system. Crucially, this concept predicts that the *confinement size*, rather than the *material of the confiner*, is the main driver of the effect.

Here, we experimentally validate this prediction by demonstrating that sub-1.5 nm silicon nanocrystals, comparable in size to previously studied gold and copper particles, can serve as effective photonic confiners, enabling broadband light emission from the surrounding bulk silicon.

Notably, we demonstrate that the observed broadband emission cannot be attributed to the conventional intrinsic emission of the nanocrystals themselves, i.e., silicon quantum dots, based on spectroscopic evidence across both frequency and time domains. In contrast, we show throughout this work that sub-1.5 nm silicon nanoparticles do not act as radiative emitters; instead, our results suggest that they function as photonic confiners that modify the local electromagnetic environment of adjacent bulk silicon domains. Importantly, the detected strong photo- and electroluminescence from silicon possesses an ultrabroad spectral width approaching 0.9 eV, and is virtually identical in spectral shape to the luminescence spectra observed when metallic confiners are used [25]. These observations support the emergence of a general radiative transition mechanism enabled by momentum-expanded photonic states. This mechanism was recently demonstrated to compensate for momentum misalignment in twisted 2D heterostructures, enhancing emission efficiency by up to four orders of magnitude, thereby confirming the effect, demonstrating its broad applicability, and remarkable transition enhancement capabilities.[26]

Our model system and its characterization are presented in Figure 1. A 300 nm-thick amorphous silicon layer was deposited onto a glass substrate to form a planar film. This film was then selectively exposed to a focused continuous-wave laser beam (30 mW, 0.75 NA, see *Methods* for details). Absorption of the laser light, accompanied by localized heating, induces crystallization of the amorphous silicon, as schematically illustrated in Figure 1a. A bright-field optical image of the laser-processed region (Figure 1b) reveals the formation of well-defined areas where the material has been modified. Two distinct regions are apparent: the light-affected zone (LAZ), located directly along the laser path, and the surrounding heat-affected zone (HAZ). As its name suggests, the HAZ forms not through direct laser exposure, but through heat and pressure diffusion originating from the LAZ.

Transmission electron microscopy (TEM) (Figure 1c) reveals the presence of sub-1.5 nm silicon nanoparticles in the LAZ. A clear increase in particle size is observed when moving outward from the LAZ, through an intermediate region, and into the HAZ, forming a spatial gradient in nanoparticle morphology (Figure 1d, see also *Supplementary Information Part I*). Importantly, electron diffraction patterns across the LAZ remain consistent, with sharp diffraction spots that indicate the presence of extended crystalline silicon domains (Figure 1e). This is further supported by Raman spectroscopy, which shows a sharp and well-defined Si–Si vibrational mode throughout the entire LAZ – a characteristic of bulk crystalline silicon with grain sizes exceeding >30 nm.[27]

In contrast, the majority of the HAZ and peripheral regions exhibit TEM images, diffraction patterns (Figure 1f), and Raman spectra that are all consistent with the presence of amorphous silicon. In sum, the model system consists of polycrystalline silicon in the LAZ, interspersed with a distribution of small sub-1.5 nm silicon nanoparticles. The size of the smaller crystallites increases progressively with distance from LAZ into HAZ, where larger particles (>5 nm) are embedded in the surrounding amorphous matrix (Figure 1c, Figure SF1).

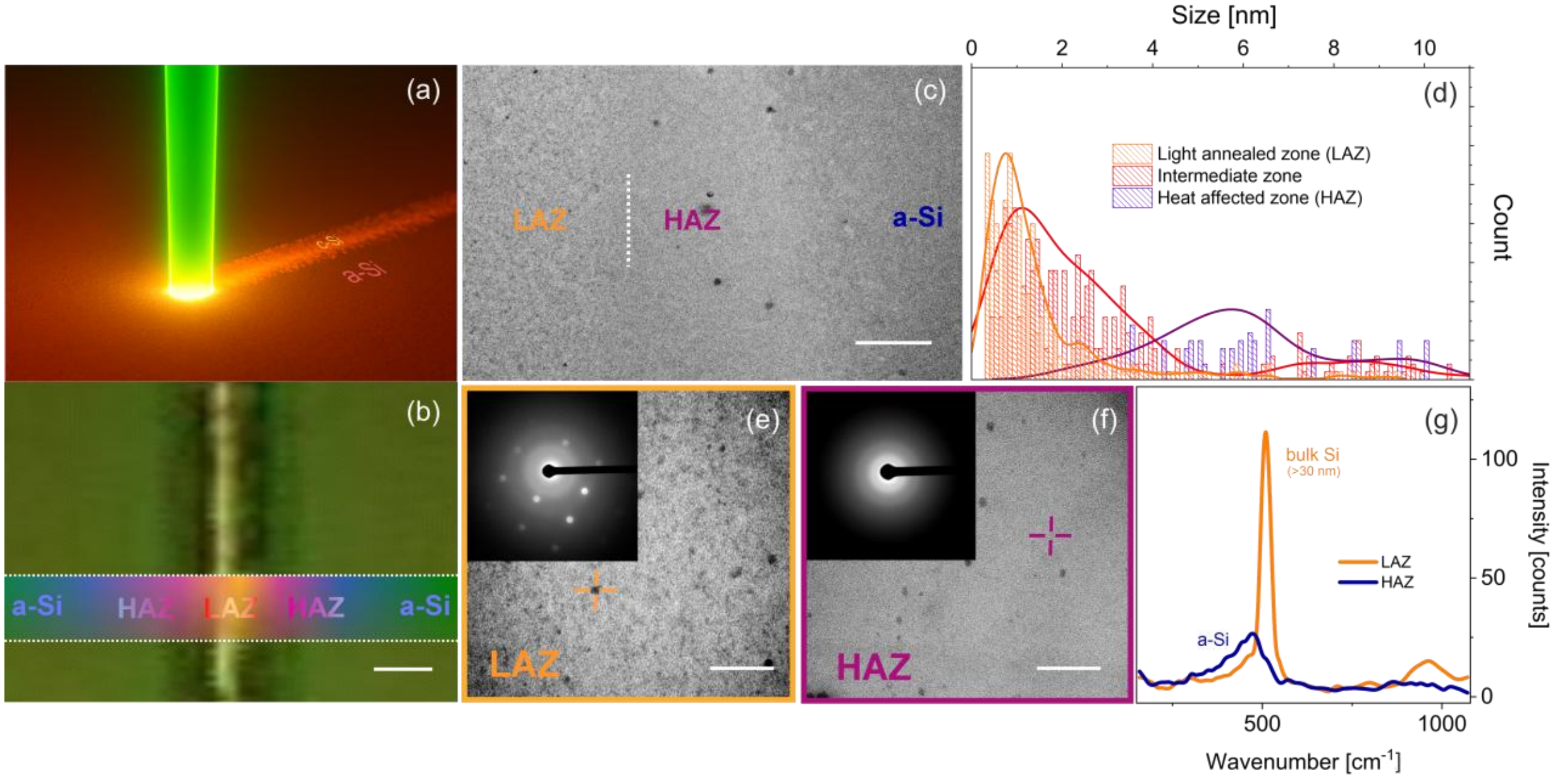

**Figure 1.** (a) Schematic representation of the exemplar silicon sample. A 300 nm-thick amorphous silicon (a-Si) layer is selectively exposed to a focused laser beam, resulting in crystallization within the light-affected zone (LAZ). The surrounding heat-affected zone (HAZ) forms not by direct laser exposure, but via thermal and pressure diffusion from the LAZ. (b) Bright-field optical image showing the LAZ, HAZ, and unmodified a-Si regions. Scale bar: 1.5 µm. (c) TEM image of the LAZ, revealing sub-1.5 nm silicon nanoparticles. Scale bar: 250 nm. (d) Nanoparticle size distribution measured across different zones, showing a rapid increase in particle size with distance from the laser-annealed region. (e, f) High-resolution TEM images of the LAZ and HAZ regions, respectively. Scale bar 75 nm. Inset: selected-area electron diffraction patterns. (g) Raman spectra acquired in the LAZ and a-Si regions. A sharp Si–Si vibrational mode in the LAZ confirms the presence of crystalline silicon domains >30 nm in size, while the broad features in the a-Si spectrum indicate its amorphous nature. Excitation: 532 nm, 0.3 mW, 0.75 NA.

Figure 2a shows the spectral emission map of the sample using photoexcitation of 0.3 mW at 532 nm and a 0.75 NA microscope objective for excitation and collection. Further excitation-dependent measurements at these powers confirm that the broadband emission observed across both LAZ and HAZ originates from photoluminescence (Figure SF2, Supplementary Information Part II). Spectral profiles taken along the LAZ-to-HAZ cross-section are shown in Figure 2b. The map and corresponding spectra reveal several important trends: a pronounced spatial dependence of the emission spectrum, both in spectral width and peak position (Figure SF3), and a rapid decrease in emission intensity as the probing spot moves into the HAZ. This behavior is in stark contrast to the characteristics expected for silicon quantum dots. We identify several independent lines of evidence that collectively and conclusively rule out a quantum dot origin for the broadband emission observed from the LAZ:

1. Quantum dots typically exhibit relatively narrow, near-Gaussian emission spectra with a bandwidth on the order of 0.2-0.35 eV (Figure SF4). For example, in the HAZ region, where 3-6 nm particles are present (Figure 1d), the emission closely resembles that of quantum dots' spectra in both symmetry and width (~0.35 eV in HAZ versus ~0.33 eV for 3–4 nm quantum dots) [16,17]. In sharp contrast, the LAZ emission spans more than 0.85 eV and exhibits a pronounced low-energy (left) shoulder, features that are atypical for quantum dot photoluminescence. While one might argue that such an ultrabroad spectrum could arise from a broad particle size distribution, this interpretation is inconsistent with the subsequent observation.
2. The observed dependence of emission efficiency is inverted relative to what is typically expected for quantum dots. The quantum yield of silicon nanocrystals drops sharply for particle sizes below 3 nm, primarily due to enhanced non-radiative recombination at the surface and trap states. [28-30] This effect is particularly pronounced in sub-2 nm particles, where the high surface-to-volume ratio strongly amplifies the role of surface defects in quenching the emission.[17] Previous studies suggest a "volcano-like" dependence of quantum yield on particle size, with an optimal size just above 3 nm, balancing strong quantum confinement with sufficient core volume to mitigate surface-related losses.[31,32] In contrast, in our experiments, the emission efficiency decreases rapidly with increasing particle size from sub-1.5 nm to 5 nm, even though particles in the 3–5 nm range would normally be expected to produce the highest brightness. This inversion of the expected size-efficiency relationship is incompatible with the conventional quantum dot interpretation.

3. The photoluminescence lifetime of ~2 ns measured from the LAZ (Figure 2e) is irreconcilable with quantum dot recombination in the observed spectral range. Silicon quantum dots emitting in the visible and near-infrared exhibit radiative lifetimes in the 1 – 100 μs range, a direct consequence of the indirect bandgap and the phonon-mediated nature of radiative recombination in quantum-confined silicon.[11,30,33,34] While nanosecond-scale emission has been reported for silicon nanostructures in the near-UV and blue spectral regions [35,36], this faster component is attributed to surface states rather than quantum confinement, and is spectrally inconsistent with the broadband visible emission observed here. Therefore, the ultrafast ~2 ns lifetimes in the visible-red to near-IR range are incompatible with a silicon quantum dot origin.

Meanwhile, the emission from the LAZ, where crystalline bulk silicon domains are surrounded by sub-1.5 nm silicon nanoparticles, exhibits an *ultrabroad* spectral profile (Figure 2d) and *ultrafast* photoluminescence lifetime (Figure 2e, τ~2 ns) nearly identical to that observed for bulk silicon wafers decorated with 1.2 nm Au or Cu nanoparticles. This striking resemblance strongly supports a common emission mechanism that is largely independent of the confiner material composition. If the emission originated from the nanoparticles themselves, significant variations would be expected due to differences in electronic structure and surface chemistry. The absence of such variations indicates that the emission arises from the silicon host material - the only common element across these three experimental systems (Figure 2d and 2e) - rather than from the nanoparticles.

While the evidence presented above rules out a quantum dot origin, it is important to consider other emission mechanisms that could, in principle, arise in a laser-processed, structurally heterogeneous silicon system. Defect-related luminescence in silicon, for example, typically produces narrow spectral bands localized at specific energies associated with known defect states, such as the ultranarrow and weak D1–D4 dislocation-related lines in the 0.8–1.0 eV range below the silicon band gap. [37,38] Oxide- and interface-related emission similarly manifests as narrow blue or green bands arising from localized Si=O or Si-OH surface states [39], which are spectrally inconsistent with the visible-to-NIR broadband emission. Stress-modified transitions, while capable of shifting emission energies locally [40], cannot account for the spatial invariance of the spectral profile observed across the LAZ on meso- and macroscopic scales. Most decisively, none of these material-specific mechanisms can explain the near-identical spectral profiles and the same ultrafast lifetime observed across three physically distinct confiner materials: gold, copper, and silicon nanoparticles, embedded in or adjacent to silicon hosts. Any mechanism tied to defects, oxides, or stress is expected to produce material- and processing-dependent spectral signatures. The absence of such variation across these three systems leaves photonic confinement and momentum expansion of the optical field as the only mechanism consistent with the full body of observations.

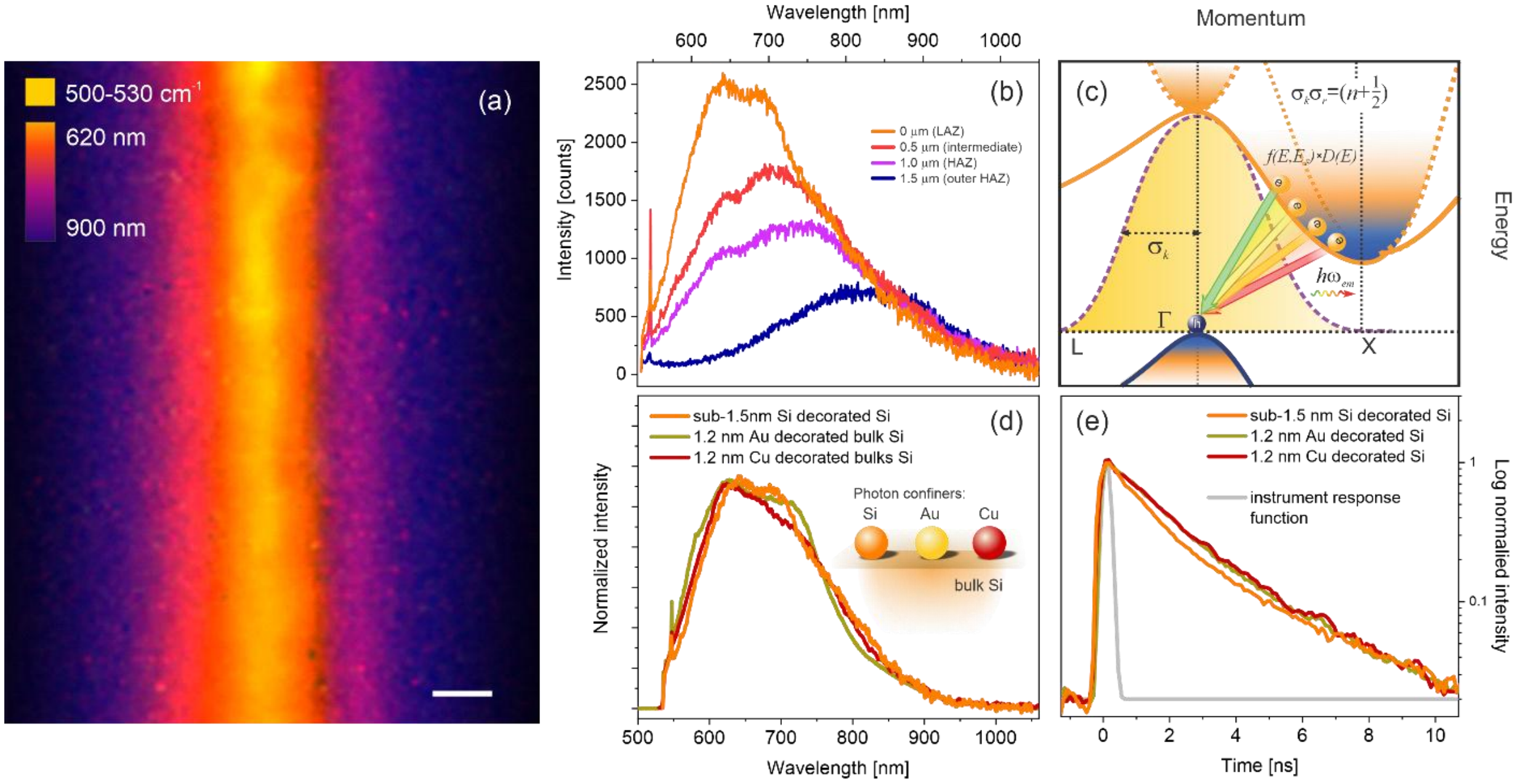

**Figure 2.** (a) Overlaid false-color Raman and PL peak position maps of the laser-annealed sample. The Si–Si Raman line (yellow) clearly indicates the presence of bulk silicon within the laser-annealed zone (LAZ). The photoluminescence in this area peaks at 620 nm (orange) and gradually shifts to longer wavelengths toward the heat-affected zone (HAZ) (transitioning into purple/blue regions, see also Figure SF3). Excitation: 532 nm, 0.3 mW, 0.75 NA. The spectral intensity decreases markedly toward the HAZ, where nanoparticle size increases significantly. (c) Schematic illustration of optical transitions. Confinement-induced momentum-expanded photonic states enhance optical absorption, producing a high-density electron gas in the conduction band. As the band becomes progressively filled, electrons occupy higher-energy states. Momentum expansion opens an efficient radiative channel for depopulating the conduction band, resulting in broadband emission. [25] (d) emission spectra and (e) photoluminescence lifetime (~2 ns) from silicon surrounded or decorated with sub-1.5 nm nanoparticles of various materials.

As previously described, confined photonic states give rise to momentum broadening (with variance $\sigma_k$), which enables strong coupling between light and electronic states in the indirect semiconductor, thereby mediating diagonal transitions, both in the absorption and emission of light by the bulk material (Figure 2c).[25,41] In the current work, the bulk silicon domains in the LAZ become efficient emitters due to the presence of adjacent sub-1.5 nm silicon nanoparticle confiners, which do not act as radiative centers, but instead impose extreme spatial confinement of the optical field, serving a role analogous to that of metallic nanoparticles in mediating the formation of momentum-expanded photonic states.

The presence of the effect is further evidenced by a careful evaluation of the emission spectra and their trends, as shown in Figure 3. High-resolution correlated maps of the photo-induced local temperature rise under continuous-wave (cw) irradiation (0.3 mW at 532 nm) and the emission peak position are presented in Figures 3a and 3b, respectively. The local temperature was determined from the spectral shifts of the bulk Si Raman line, which is highly sensitive to temperature changes. Representative Raman spectra and the calibration curve for the Raman peak position as a function of temperature are provided in Supplementary Information Part III (Figures SF5 and SF6).

The maps reveal a clear correlation: the hottest regions within the LAZ (Figure 3a) coincide with the most blue-shifted emission (Figure 3b). This indicates that large bulk crystalline structures within the LAZ reach significantly elevated temperatures under focused cw illumination due to enhanced absorption facilitated by momentum expansion on sub-1.5 nm particles. This observation is consistent with previous studies showing that momentum expansion can increase optical absorption by up to three orders of magnitude.[41] As the particle size increases across the LAZ and toward the HAZ, their larger spatial extent can no longer provide the momentum expansion necessary for efficient diagonal transitions. Consequently, the temperature drops markedly, as absorption reverts to values typical for bulk (indirect) silicon.

These observations also suggest that momentum expansion in small particles plays a major role in the initial particle formation from the surrounding amorphous matrix. Figures 3c and 3d show the evolution of temperature and emission spectra center of mass, respectively, measured from a single spot in the LAZ subjected to prolonged and repeated laser exposure (see also Figure SF7, Supplementary Information Part IV). Both parameters display correlated saturation behavior.

These data indicate that initially, the enhanced absorption rate, due to photon momentum expansion by sub-1.5 nm particles, enables the surrounding material to reach high temperatures, promoting efficient crystallization. Once the particle size approaches ~3 nm, the momentum expansion effect subsides, and the enhanced absorption mechanism switches off, marked by the stabilization of both temperature and peak position.

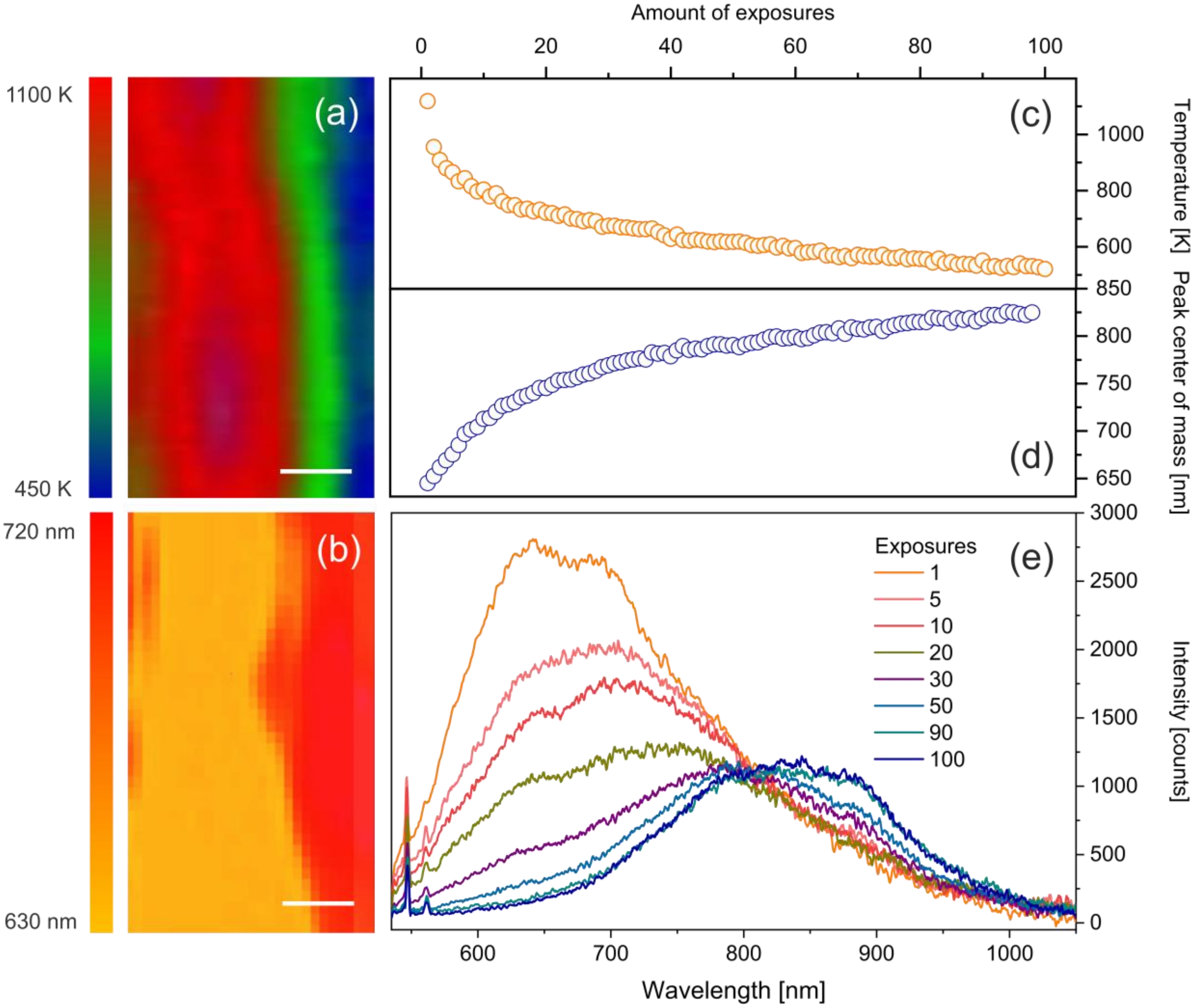

**Figure 3.** (a) Map of the photo-induced temperature rise in the LAZ derived from the spectral peak position of bulk silicon's optical phonon Raman line (see Supplementary Information Part III). Excitation: 532 nm, 0.3 mW, 0.75 NA. Scale bar: 750 nm. (b) Emission map of the same LAZ region. Scale bar: 500 nm. Evolution of (c) temperature and (d) emission spectral center-of-mass with the number of light exposures. Each light exposure consisted of 0.3 mW for 10 s. (e) Evolution of the emission spectrum with increasing laser exposures. The presence of an isosbestic point strongly supports a transition between two distinct spectral components, indicating different dominant emission mechanisms in each regime.

Strong support of this notion can be found in the spectral trend shown in Figure 3e. Remarkably, the spectral evolution under sustained laser illumination exhibits a clear isosbestic point, indicating a transition from one dominant emission mechanism to another. Initially, sub-1.5 nm particles, poor light emitters on their own, act as light confiners, providing the momentum expansion required for broadband emission from large bulk silicon structures in the LAZ (orange intense spectrum). As these particles grow during light exposure, the emission pathway from bulk silicon domains in the LAZ is progressively suppressed. Concurrently, the particles themselves approach the previously discussed "volcano-like" optimal size, producing a narrow emission band centered at 800–850 nm, all consistent with >3 nm (dark blue spectrum) particles and in excellent agreement with literature.[17,29,31] The persistence of the isosbestic point is inconsistent with a continuous size distribution of emitters and, instead, indicates the coexistence of these two distinct emission pathways: broadband, momentum-driven emission from bulk silicon enabled by sub-1.5 nm light confiners at early exposures, and narrower, quantum-dot-like emission from optimally sized (>3 nm) grown nanocrystals that emerge at later exposures.

We next use this silicon model system to achieve electrically driven luminescence (EL) from the LAZ region (Figure 4). For these experiments, the amorphous silicon was deposited on a glass substrate, with two copper electrodes patterned on opposite sides of the device, as schematically shown in Figure 4a. The LAZ was written across the two electrodes using the same laser-annealing procedure described above (Figure 4b; see *Methods* and *Supplementary Information Part V* for fabrication details and parameters). A bright-field optical image of the selected device area is shown in Figure 4c. Upon application of a 15 V bias, a bright visible emission emerges. EL spectral mapping (650–700 nm) shown in Figure 4d confirms that the emission originates exclusively from the LAZ. As expected, the emission requires moderately high voltages, as is usual for the case of undoped semiconductors. In low or undoped semiconductors, in the absence of sufficient carrier densities, electroluminescence can only arise from high-field-assisted tunneling and impact ionization by accelerated hot carriers, requiring a substantial bias voltage.[42-44] The emission from the LAZ is sufficiently intense to remain clearly visible by the microscope's low-efficiency camera (see *Methods* and Supplementary Video 1 and Video 2, Supplementary Information Part VI).

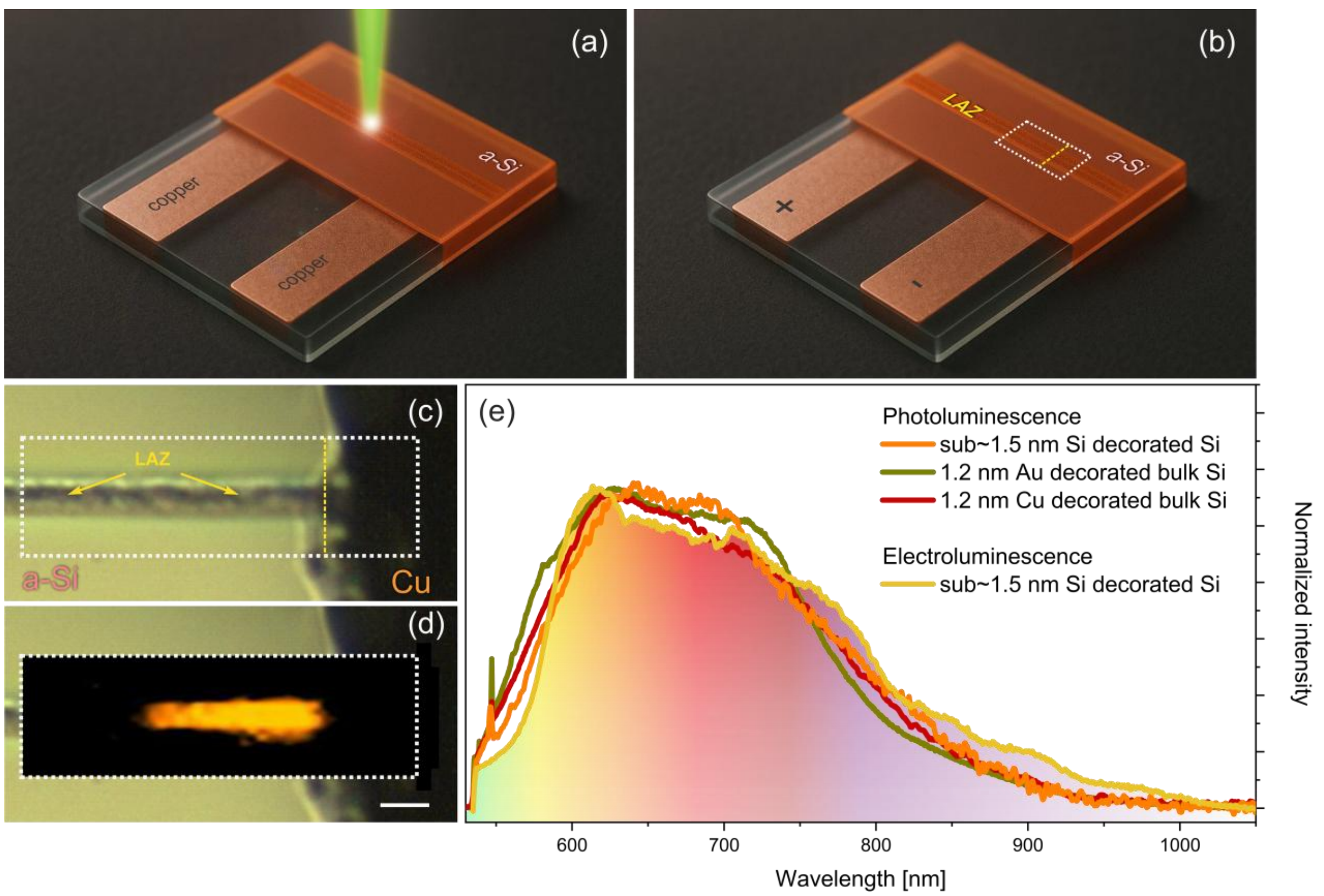

**Figure 4.** (a) Schematic representation of the device used for electroluminescence (EL) measurements (see *Supplementary Information Part V*). (b) Light-affected zone (LAZ) formed in amorphous silicon deposited on glass with copper electrodes patterned on the sides of the device. (c) Bright-field image and (d) EL emission spectral map acquired in the 650–700 nm range at 15 V applied voltage. Scale bar 1 μm. (e) EL spectrum (yellow, 15 V) compared with photoluminescence (PL) from the LAZ (orange) and PL from bulk silicon decorated with 1.2 nm Au (dark yellow) and 1.2 nm Cu (red) nanoparticles. The strong spectral agreement supports a common physical origin: momentum-expanded photonic states enable direct optical transitions from high-energy conduction-band states to the valence band, bypassing the phonon-mediated transitions typical of indirect semiconductors.

Ultimately, as shown in Figure 4e, a close inspection of the EL spectrum reveals that it matches the PL spectrum from the same region, along with the emission from bulk silicon decorated with 1.2 nm metal nanoparticles. *This strong spectral agreement supports a common origin of the underlying physical mechanism*: momentum-expanded photonic states enable direct optical transitions from high-energy conduction-band states to the valence band, bypassing the phonon-mediated processes typical of indirect semiconductors. As in previous PL studies, when local carrier densities in the conduction band exceed $\sim 10^{20}$ $cm^{-3}$ ($\sim$ 0.1 electron/$nm^3$), electrons populate progressively higher-energy states, maximizing spectral overlap between the electron distribution and the transition probability $P(\sigma_k)$, defined as the projection of the Gaussian photon momentum distribution onto the conduction-band dispersion profile.[25] Following this notion, a higher input voltage increases the density of hot energetic electrons in the conduction band, injected into the bulk structures within the LAZ via tunneling under the applied DC field. Consistent with this picture, we observe that EL spectra can exhibit a blue shift of the first moment of the spectral profile with increasing input voltage (Figures SF14), while the red side of the emission remains nearly unchanged - a signature of the elevated Fermi level driven by the high electron density in the conduction band, as discussed in previous work [25]. Importantly, across all EL sites studied, the spectra remain consistent with the photoluminescence shape originating from momentum-expanded photonic states, with no evidence of electroluminescence from > 3 nm quantum dots.

Together, these results uncover a previously unrecognized effect that enables efficient radiative recombination and broadband photo- and electroluminescence from silicon. Crucially, the data show that the observed emission does not originate from nanoscale silicon emitters themselves, but from bulk-like silicon domains whose radiative pathways are fundamentally altered by extreme photonic confinement. This distinction separates the present mechanism from conventional quantum confinement and establishes a qualitatively different route to light emission in indirect semiconductors. The emission spans hundreds of meV and vanishes once sub-2 nm spatial confinement is lost, confirming the size-dependent nature of the mechanism.

We note that the present work represents a qualitative demonstration of a new physical mechanism rather than an attempt to quantitatively optimize device performance. The absolute quantum yield of the current system is low, estimated at 0.2 %, as expected for an unoptimized proof-of-concept device architecture based on a single undoped semiconductor layer under high voltage conditions (see Supplementary Information Part VIII). However, this is not an intrinsic limitation of the mechanism itself. Several clear routes toward efficiency improvement exist: implementing a proper diode architecture through the introduction of a pn junction, optimizing the confiner size distribution and density within the junction, to name just a few. The remarkably high quantum efficiency of photoexcited emission recently demonstrated in bulk silicon using metal confiners [25], as well as in 2D heterostructures via the same photonic momentum expansion mechanism [26], suggests a high ceiling for efficiency improvement in the present system. Systematic optimization of these parameters is the subject of ongoing work.

In direct-gap semiconductors, conduction-band electrons, whether photoexcited or electrically driven, quickly relax to the conduction-band minimum, where rapid radiative recombination depletes carriers and yields narrow, band-edge emission. Silicon behaves fundamentally differently. Its indirect bandgap, long regarded as a limitation for optoelectronic applications, becomes a powerful advantage when combined with momentum-expanded photonic states induced by extreme spatial confinement. Instead of being funneled into a narrow emission channel, carriers accumulate in higher-energy conduction-band states, where the transition probability $P(\sigma_k)$ is maximized, opening radiative pathways that bypass phonon bottlenecks. This results in intense, broadband emission from a material traditionally considered optically inactive.

Building on these results, we identify the onset of a new class of light–matter hybrid states, where the Heisenberg uncertainty principle directly links spatial confinement of light with its reciprocal-space expansion. In this framework, silicon combined with properly prepared confined light fields forms an emergent *photonic Heisenberg matter* - a hybrid regime in which strongly confined photons reshape the landscape of allowed electronic transitions, unlocking absorption and emission pathways inaccessible in conventional photonic systems.

While direct-bandgap materials struggle to produce intrinsically broadband electroluminescence, our findings show that it is precisely silicon's indirect bandgap, when coupled with photon confinement and Heisenberg-driven momentum expansion, that uniquely facilitates the generation of broadband emission across the visible spectrum. With further optimization, this discovery opens a pathway toward a new generation of silicon-based optoelectronic devices, including general-purpose broadband LEDs, ultracompact emitters for integrated photonic circuits, and, ultimately, all-silicon coherent light sources, provided that gain mechanisms and competing non-radiative pathways such as Auger recombination can be further characterized and optimized.

## Methods

*Amorphous silicon deposition*

Amorphous silicon (a-Si) thin films were deposited onto borosilicate microscopy cover slips (170 μm thickness) using a modified Gatan Precision Ion Polishing System (PIPS, Gatan Inc.) operated as a sputter coater. The procedure closely followed previously reported methods[45], with modifications enabling both coating of bulk sample surfaces and deposition of thin films from various materials. An advantage of using the PIPS for this purpose is its oil-free vacuum system, which minimizes carbon contamination during deposition. The sputter source consisted of a ~5 mm diameter monocrystalline wafer mounted on the standard specimen holder post. The milling system incorporates two Penning-type ion guns (PIGs), operated simultaneously to accelerate the sputtering process. The angle between the guns and the rotating target material was fixed at +15° throughout the deposition. The system was operated at a maximum accelerating voltage of 8 kV. Prior to deposition, the wafer source was sputter-cleaned for 2 minutes with the pneumatic shutter closed. The base vacuum before deposition was typically $\leq 1 \times 10^{-3}$ Pa. Typical sputtering times ranged from 30 to 90 minutes, depending on the sputter source and the desired film thickness. The resulting films had thicknesses of 150–450 nm and were uniform across the substrate.

*Device fabrication for electroluminescence study*

To fabricate a thin, non-crystalline a-Si layer between conductive contacts, a copper target was thermally evaporated onto borosilicate coverslips to form a uniform conductive coating. Thermal evaporation was performed under high vacuum ($\leq 1 \times 10^{-4}$ Pa) to ensure strong adhesion and minimize contamination. The conductive layer was then mechanically divided into two electrically isolated regions by scratching it with a sharp tungsten cantilever under an optical microscope, creating a 7–15 μm gap (see *Supplementary Information Part V*, Figure SF8). This gap size was selected to balance low contact resistance with the spatial resolution required for localized electrical measurements across the a-Si layer. Following gap formation, a-Si was deposited onto the Cu contacts through a mask with a 1 mm circular aperture positioned directly over the gap. The aperture size defined the active device area and ensured reproducible alignment with the contact gap. The mask was held in close proximity to the substrate to minimize lateral deposition and produce sharp feature edges. This fabrication method enabled the formation of a well-defined a-Si layer suitable for subsequent electrical and structural characterization.

*Device electrical characterizations and current/voltage supply*

The Metrohm Autolab potentiostat–galvanostat (Metrohm AG, Barendrecht, Netherlands) was used to perform I–V curve measurements. Devices were connected to the analyzer via the Autolab PGSTAT electrode set (Metrohm, Switzerland). Measurements were controlled using NOVA software (Metrohm, Switzerland), with calibration of PGSTAT cell factors performed before data acquisition.

A DC power supply (Keithley 2400, Keithley Instruments, USA) was used to apply either pulsed or constant voltages to the sample in the range of 0–20 V. Electrical resistance variations and current kinetics during testing were recorded using the Keithley 2400 configured as a digital multimeter and automated via a GPIB interface.

*Light-induced silicon crystallization*

For light-assisted crystallization of silicon, a 30 mW continuous-wave (cw) laser at 532 nm was used. The beam was focused onto the sample surface using a 50×, 0.75 NA objective lens. To create a prolonged linear light-affected zone (LAZ), the sample stage was translated under the objective using an automated piezo stage at a linear speed of 500 μm/s. This scanning speed was selected to produce a uniform crystallization effect and consistent temperature distribution along the scan direction. All procedures were conducted in an argon atmosphere (oxygen content < 0.1 %) to minimize oxidation.

*Raman and PL micro-spectroscopy and mapping*

All micro-spectroscopy experiments were performed on a custom-modified microscopy system based on an InVia Renishaw system using 532 nm and 785 nm laser sources. The samples were illuminated with a 0.75 NA air objective (Leica), and emission was collected in the epi-configuration. The spectra were measured using both standard and expanded modes with a 1200 gr/mm and 2400 gr/mm diffraction gratings used for spectral scanning with 0.1 $cm^{-1}$ resolution over the entire spectral region.

*Transmission electron microscopy*

Transmission electron microscopy (TEM) imaging and electron diffraction at the nanoscale were performed using a Hitachi HT7700 Exalens microscope (Hitachi High-Tech Science Corporation, Japan) operated at an accelerating voltage of 120 kV, providing a spatial resolution of 0.144 nm. Images were captured with an AMT XR-81 CCD camera (2742 × 3296 pixels, 8 megapixels, 5.5 μm pixel size), enabling high-quality structural characterization. For TEM sample preparation, the amorphous silicon (a-Si) film was deposited onto copper grids coated with a carbon support film (EMresolution, Sheffield, UK). Laser-assisted crystallization and annealing were carried out using the standard parameters described throughout the manuscript. TEM images and diffraction patterns were processed and analyzed using Fiji/ImageJ 2.1 to extract quantitative information on nanoparticle size and lattice spacing.

*Time-resolved fluorescence measurements*

Time-resolved photoluminescence measurements have been conducted using Leica SP8 Falcon system, a customized commercial imaging platform, equipped with white-light picosecond laser source, which consists of a high-energy pulsed IR-fiber laser that is fed through the photonics crystal fiber to generate a spectral continuum. The fluorescence lifetime expeirments were conducted in epi-geometry using 10x objective, 514 nm excitation and signals detected using FALCON FLIM detector (Leica). Time resolution 120 ps.

*Videography*

Microscale electroluminescence videos were captured in a wide-field configuration using a standard back-illuminated CMOS camera (752x480). The camera is native to the microscopy system used for sample orientation in bright-field imaging.

**Data availability**

The data is available from the corresponding author on reasonable request.

**Competing interest**

The authors declare no competing interests.

**Acknowledgements**

D.A.F. thanks Yulia Davydova for support and help during this study. The authors thank Prof. V Ara Apkarian, Prof. Sasha Chernyshev, Prof. Paul H.M. van Loosdrecht, Prof. Maxx Arguilla for fruitful discussions. D.A.F. and E.O.P acknowledge funding from Chan Zuckerberg Initiative 2023-321174 (5022) GB-1585590, NSF 2434622, and NIH S10-OD028698. D.A.F. dedicates this work to beloved friend Fiona.